\documentstyle[pre,aps,multicol]{revtex}
\begin{document}
\draft
\title{Asymptotic distributions
of periodically driven stochastic systems}
\author{Sreedhar B. Dutta and Mustansir Barma}
\address{Department of Theoretical Physics, Tata Institute of Fundamental
Research, Homi Bhabha Road, Mumbai 400 005, India}
\date{2 January 2003}

\maketitle
\begin{abstract}
We study the large-time behaviour of Brownian particles moving through a
viscous medium in a confined potential, and which are further
subjected to position-dependent driving forces that are periodic in time.
We focus on the case where these driving forces are rapidly
oscillating with an amplitude that is not necessarily small. We develop a
perturbative method for the high-frequency regime to find the large-time
behaviour of periodically driven stochastic systems. The asymptotic
distribution of Brownian particles is then determined to second order.
To first order, these particles are found to execute small-amplitude
oscillations around an effective static potential that can have
interesting forms.
\end{abstract}

\begin{multicols}{2}
\section{Introduction}
\label{intro}
When a macroscopic system characterised by a Hamiltonian $H(C)$ is in
contact with an environment with temperature $\beta^{-1}$, at long
enough times it reaches an equilibrium state and is described by the
Boltzmann-Gibbs distribution $P_{eq}(C) \sim \exp[-\beta H(C)]$ 
over the configurations of the system.  If, however, the
system is subjected to time-dependent forces of appreciable
magnitude, there is no analogous general statement that can
be made about the distribution $P(C,t)$ at large times.  A case of
particular interest arises when these forces vary periodically in time.  
While it is straightforward to see that the
large-time distribution must be periodic as well, its full form is not
known in general.  It is thus of interest to seek explicit answers
for particular physical systems subjected to periodic driving.
 
In this paper, we focus on a paradigmatic system: a Brownian
particle which feels viscous forces and random impulses from the
surrounding medium, and is confined by a potential well.  We ask: What
is the effect of a further oscillating potential on the state of the
particle?  We are primarily interested in the case where the
fluctuating potential has a nontrivial spatial dependence.
We analyze the problem mostly in the high-frequency limit 
and find the asymptotic state perturbatively. The resulting time-averaged
asymptotic state is described effectively by a distribution of the
Boltzmann form with an energy function that has three parts: 
a kinetic energy term that depends only on velocity; an effective
potential that depends only on position coordinates; and a term with both
velocity and position coordinates. However, to the leading order the
result is particularly simple, and involves only the kinetic term and an
effective frequency-dependent potential energy whose form can be specified
exactly. 

The effects of rapidly oscillating periodic forces on purely mechanical
systems were demonstrated many years ago for a driven pendulum
\cite{kapitza} and were also studied for more general cases \cite{landau}.
Periodically driven stochastic systems too have been studied extensively
over the past two decades in the context of stochastic resonance 
(see, {\it e.g.}, Ref.\cite{gammaitoni}, and references therein). 
However, most of these studies were restricted to weak periodic driving
with position-independent forces\cite{devoret,jung,presilla,hu}, often
only in the overdamped regime. Further, it has sometimes been assumed
that the driving frequency is larger than all typical frequencies,
including that which is associated with the noise \cite{jung}. 
Our treatment generalizes that of Refs. \cite{devoret} and \cite{jung} in
the high-frequency regime, by allowing or arbitrary damping; by including
position dependence of the driving force; and by taking frequencies to be
higher than those set by the confining potential, yet not necessarily
higher than those set by the noise.

Some applications of our results are possible.  For instance,
depending on the spatial variation of the fluctuating forces, the
effective potential may have more than one local minimum even if the
original potential has only one. Under such circumstances, an assembly of
Brownian particles would tend to segregate in two separate collections.
Moreover, the fact that the effective potential depends on physical
properties, such as the mass of the particles, can be exploited to promote
segregation of two sets of Brownian particles that differ from each other
in mass or some other physical attribute.  

The layout of the paper is as follows.  
In the following section, 
we define the problem and discuss the different time scales involved and
their interplay, and discuss various regimes qualitatively.  
In Sec.  \ref{formalism}, 
we develop the necessary formalism required to
address rapid periodic drive and arrive at a perturbative scheme.
In Sec. \ref{asympert}, 
we use this scheme to determine the asymptotic distribution.
In Sec. \ref{slowfields}, 
we briefly discuss the effect of slow periodic driving so as to compare
with that of rapid driving. Finally, in Sec. \ref{concl},
we conclude with a discussion of possible directions in which our
results may be generalized, possible applications (particle
segregation, particle sifting), and an example (the simple pendulum) that
demonstrates the significance of the new effects found.

\section{Periodically driven Brownian Particle}
\label{pdbp}
   We shall consider a one-dimensional Brownian particle in a potential
well, subjected to a periodic force along with a damping force and 
random noise. The equation of motion of the driven Brownian particle
moving in a viscous environment is  
\begin{equation}
\label{bpeom}
 m\ddot{x}~=~ -\gamma \dot{x} ~-~ \frac{\partial}{\partial x} U(x) 
~+~ F(x,t)~+~ \eta(t),   
\end{equation}
where $m$ is the mass, $\gamma$ is the coefficient of viscosity, 
$U(x)$ is a static confining potential, $F(x,t)$
is the periodic driving force with a period $T$, $F(x,t)=F(x,t+T)$, and
$\eta(t)$ is a Gaussian random noise with 
$\langle \eta(t)\rangle_{\eta} = 0$ and 
$\langle \eta(t)\eta(t')\rangle_{\eta} = 2\gamma \beta^{-1}~\delta(t-t')$, 
where $\beta^{-1}$ is the temperature of the surrounding heat bath. 

 The probability distribution of the Brownian particle is defined as 
$
 P(x,v,t)= 
\langle\delta(x-x_{\eta}(t))\delta(v-\dot{x}_{\eta}(t))\rangle_{\eta}
$,
where $x_{\eta}(t)$ and $\dot{x}_{\eta}(t)$ are the
position and the velocity at time $t$ for a particular history of
$\{\eta(t)\}$ over a time $t$. The time evolution of $P(x,v,t)$ is
described
by the following Fokker-Planck (FP) equation (also
referred to as the Kramers equation),    
\begin{eqnarray}
\label{FPeqn}
\frac{\partial P}{\partial t} =
&-&\frac{\partial}{\partial x}(vP)
-\frac{\partial}{\partial v} \biggl[ \frac{1}{m}
\{-\gamma v -  U'(x)  + F(x,t)\} P
\biggr] \nonumber \\ 
&+& \frac{\gamma}{\beta m^2}\frac{\partial^2 P}{\partial v^2}~.
\end{eqnarray}
The driving force $F(x,t)$ that is oscillating with a frequency
$\omega = 2\pi/T$ is chosen to be 
\begin{equation}
\label{forceeqn}
F(x,t)~=~ f(x)~\cos(\omega t) ~+~ g(x)~\sin(\omega t)
\end{equation}
in the domain $L_1 < x < L_2$, where the amplitude functions $f(x)$ and
$g(x)$ vanish at the boundaries, $x=L_1$ and $L_2$, and outside the
domain. This choice of $F(x,t)$ is made 
for convenience; choosing a more general periodic function will
not hinder our analysis. The generalization to higher dimensions and to 
many interacting particles is also straightforward. 
Our aim is to find the large-time distribution
that we denote by
$P_{\infty}(x,v,t) =  \lim_{t \rightarrow \infty} P(x,v,t)$. 

In the absence of a driving force, all solutions $P(x,v,t)$
of the FP equation, corresponding to various arbitrary initial
distributions, tend to a unique distribution after
a long enough time\cite{riskenkampen}.
This distribution $P_{\infty}(x,v,t)$ for the Brownian particles 
takes the equilibrium canonical form
$P_{eq}(x,v) = (1/Z_0)\exp\left\{-\beta \left[\frac{1}{2}mv^2 +
U(x)\right]\right\}$. 
When a periodic driving force is present, $P_{\infty}(x,v,t)$ approaches a
periodic function of time which is unique up to a 
phase\cite{jungreview,gammaitoni}. In brief, the argument for the
periodicity goes as follows. When the FP operator is periodic,
${\cal L}(t)={\cal L}(t+T)$, the solution to the FP equation 
$[\partial_t - {\cal L}(t)]P(x,v,t)=0$  can be expanded in terms of the
Floquet-type functions $p_{\mu}(x,v,t)\exp(-\mu t)$. The functions 
$p_{\mu}(x,v,t)=p_{\mu}(x,v,t+T)$ are periodic and are the right
eigenfunctions of $\partial_t - {\cal L}(t)$ with eigenvalues
$\mu$. It is known that for an $N$-dimensional FP equation, the
real parts of these eigenvalues Re$(\mu)$ are
positive semidefinite and hence in the large-time limit, for typical
confining potentials, only $p_{0}(x,v,t)$ survives.
We are interested in finding this large-time distribution 
$P_{\infty}(x,v,t)$ for a given $F(x,t)$.
Since no analytic solution of the FP equation is known for an
arbitrary time periodic $F(x,t)$, even when it involves only the
fundamental frequency, we shall restrict our attention to
certain regimes of the driving frequency while solving for 
$P_{\infty}(x,v,t)$.

In the absence of the driving force there are two important time scales in
the system; one, $\tau_v = m / \gamma$, is introduced by the viscous
medium and the other, $\tau_w= 2\pi / \omega_0$, is related typically to
the curvature at the bottom of the potential well,
$\omega_0=\sqrt{U''(x_{min})/m }$.   
The velocity variable equilibriates in a time scale set by $\tau_v$ and
hence for larger times it gets described
by a stationary distribution. Thus, in a highly viscous medium, where
$\tau_v \ll \tau_w$ holds, after a time $t \gg \tau_v$ the distribution
can be written as $P(x,v,t)= P_{eq}(v)P(x,t)$, where 
$P_{eq}(v) \sim \exp\left[ -\beta \frac{1}{2}m v^2\right]$
is the canonical distribution for velocity and
$P(x,t)=\int dv P(x,v,t)$ is
the marginal distribution involving only position. 
This marginal distribution satisfies an FP equation (often called the
Smoluchowski equation) obtained by dropping the inertial term
in Eq.(\ref{bpeom}), as this term becomes insignificant in comparison
with the viscous term once the time exceeds $\tau_v$.

If, however, there is a driving force with a high enough frequency then
this reduction to the Smoluchowski limit does not take place.
The oscillating driving force introduces one other important time scale
associated with the timeperiod $T$ whose existence
restricts the domain of validity of the 
Smoluchowski equation even in the high-friction limit.
It is useful to demarcate different regimes of the driving frequency 
in this limit:
$(a)~ \tau_v \ll \tau_w < T$ and $(b)~ \tau_v < T \ll \tau_w$ or
$ T < \tau_v \ll \tau_w$. 
In case $(a)$, the velocity decouples from the
position and one can still use the Smoluchowski equation to obtain the
large-time distribution, while in case $(b)$, the driving force
has a time scale comparable to that of the velocity relaxation time and
hence it is necessary to retain the Kramers equation.

If the time scale of
measurement exceeds the time period $T$, then the relevant quantity
is the  large-time distribution averaged over a time period,  
$\overline{P_{\infty}(x,v,t)}=(1/T)\int_{0}^{T} P_{\infty}(x,v,t)$. 
Hence we shall also determine $\overline{P_{\infty}(x,v,t)}$. 

\section{Rapidly Oscillating forces: Formalism}
\label{formalism}
  In a mechanical system, in the absence of a viscous force and random
noise, it is known that for $\omega \gg \omega_0$, the particle executes
small amplitude oscillations of frequency $\omega$ about a smooth mean
path \cite{landau}. This motion can be described by separating it into
slow and fast variables; the fast variable decouples from the slow one
while the slow variable sees a static effective potential, modified due
to the oscillations of the fast variable. 
In this section, we develop the necessary formalism that accommodates 
an analogy with this separation into slow and fast variables and enables
us to solve the problem when viscous and random forces
are included. 

\subsection{Transformation of FP
equation}
\label{trans}
In this subsection, we transform the FP equation under a specific
coordinate transformation which enables it to be solved perturbatively.
For the perturbative treatment to be valid a sufficient condition, though
not necessary, is that $\omega$ be large, while no assumption is made
about the amplitude of the driving force in comparison with the static
potential.

We make the coordinate transformation $\{x,v,t\} \rightarrow
\{X,V,\tau\}$ under which the distribution is made to behave
like a scalar function:
$P(x,v,t) \rightarrow \tilde{P}(X,V,\tau) = P(x,v,t)$.
The old and new coordinates are related as follows,
\begin{equation}
\label{coord}
x = X + \xi(X,\tau), 
~v = V +  \frac{\partial }{\partial \tau} \xi(X,\tau),
~t = \tau.
\end{equation}
The explicit form of $\xi(X,\tau)$ will be specified later. Note that
the volume element $dxdv = dXdV J(X,V,\tau)$, where
$J(X,V,\tau)= (1+ \partial \xi /\partial X )$ and hence, 
if $P$ is the probability density in $(x,v)$ space then
$\tilde{P}J$ becomes the the probability density in $(X,V)$ space.
Also, under the above coordinate transformation the derivatives transform
as follows: 
\begin{eqnarray}
\label{deriv}
\frac{\partial }{\partial x}
&=& \frac{1}{1+\xi'} ~\frac{\partial }{\partial X}
- \frac{\dot\xi'}{1+\xi'} ~\frac{\partial }{\partial V}, \nonumber \\
\frac{\partial }{\partial v}
&=& \frac{\partial }{\partial V}, \nonumber \\
\frac{\partial }{\partial t} &=&
\frac{\partial }{\partial \tau}
-\frac{\dot\xi}{1+\xi'}~ \frac{\partial }{\partial X}
+\left( \frac{\dot\xi \dot\xi'}{1+\xi'} - \ddot\xi
\right)~\frac{\partial }{\partial V}, 
\end{eqnarray}
where the {\it dots} and {\it primes} on $\xi$ stand for derivatives with
respect to $\tau$ and $X$, respectively. Making use of Eqs. 
(\ref{coord}) and (\ref{deriv}) in Eq.(\ref{FPeqn}), we get
\begin{eqnarray}
\label{NFPeqn}
\frac{\partial \tilde{P}}{\partial \tau} =
&-&\frac{\partial}{\partial X}(V\tilde{P})
-\frac{\partial}{\partial V} \biggl[ \frac{1}{m}
\{-\gamma V +f_{U}(X+\xi)  \nonumber \\
&+& 
F(X+\xi,\tau)-F(X,\tau) \} \tilde{P} \biggr] 
+  \frac{\gamma}{\beta m^2} 
\frac{\partial^2 \tilde{P}}{\partial V^2}
\nonumber \\
&+& V\biggl[
\frac{\xi'}{1+\xi'}\frac{\partial \tilde{P}}{\partial X}
+\frac{\dot\xi'}{1+\xi'} \frac{\partial \tilde{P}}{\partial V}
\biggr]  \nonumber \\
&+& \frac{1}{m}\biggl[ m\ddot\xi + \gamma \dot\xi - F(X,\tau)\biggr]
\frac{\partial \tilde{P}}{\partial V},
\end{eqnarray}
where $f_{U}(X)= -\partial U(X)/\partial X$.   

Now choose $\xi(X,\tau)$ such that it is a solution of the following
equation
\begin{equation}
\label{xieqn}
m\ddot\xi = -\gamma \dot\xi + F(X,\tau)~.
\end{equation}
For $F(X,\tau)$ as given in Eq. (\ref{forceeqn}), the solution to
the above equation is  
\begin{eqnarray}
\label{xisoln}
\xi(X,\tau)&=& \frac{-1}{m(\omega^2 + \frac{\gamma^2}{m^2})}
\biggl[ \biggl( f(X)+\frac{\gamma}{m\omega}g(X) \biggr)\cos(\omega \tau)
\nonumber \\
& &+ 
\biggl( g(X)-\frac{\gamma}{m\omega}f(X) \biggr) \sin(\omega\tau)
\biggr].
\end{eqnarray}
Since $\xi$ is small for large values of 
$(\omega^2 + \gamma^2/m^2)$, we may expand Eq.(\ref{NFPeqn})
perturbatively in $\xi$. 
The reason for choosing the above dynamics (Eq.(\ref{xieqn})) for
$\xi$ is that the last  term in Eq.(\ref{NFPeqn}) becomes zero
and further the explicitly (time)$\tau$-dependent term becomes small if
$\xi$ is small, thus making the equation amenable to perturbative
analysis. 

Note that when the driving force is $x$ independent Eq.(\ref{NFPeqn})
reduces to the usual FP equation with the static force $f_U(X)$
being replaced by $f_U(X+\xi)$.

\subsection{The perturbative scheme}   
\label{Pert}
We now find the large-time solution of Eq.(\ref{NFPeqn}) 
perturbatively in powers of $\xi$. Upon substituting $\xi$ as given in
Eq.(\ref{xisoln}), we see that Eq.(\ref{NFPeqn}) takes the form
\begin{eqnarray}
\label{pertFP}
\frac{\partial }{\partial \tau} \tilde{P}(X,V,\tau)
&=& [ {\cal L}+ \Delta {\cal L} ]\tilde{P}(X,V,\tau)
\nonumber \\ 
&\equiv&  \sum_{n=0}^{\infty} 
[{\cal L}^{(n)}+\Delta {\cal L}^{(n)} ] \tilde{P}(X,V,\tau),
\end{eqnarray}
where ${\cal L}$ and $\Delta {\cal L}$ are static and time-dependent
operators, respectively.
The superscript on the operators ${\cal L}^{(n)}$ and 
$\Delta {\cal L}^{(n)}$
indicates that they are of $O(\xi^n)$; the explicit forms of
the first few operators are  
\begin{eqnarray}
{\cal L}^{(0)} &=&
- V \frac{\partial}{\partial X}
+\frac{1}{m}\frac{\partial}{\partial V} 
\left[ \gamma V + U'(X) \right] 
~+~ \frac{\gamma}{\beta m^2} \frac{\partial^2 }{\partial V^2},
\nonumber \\
{\cal L}^{(1)} &=&
-\frac{1}{m} \overline{\xi F'(X,\tau)} 
~\frac{\partial}{\partial V}, 
\nonumber \\
{\cal L}^{(2)} &=&
\frac{1}{2m}\overline{\xi^2} U'''(X) 
\frac{\partial}{\partial V} ~-~ 
V \overline{\xi'^2} \frac{\partial}{\partial X},
\nonumber \\
\Delta {\cal L}^{(0)} &=& 0,
\nonumber \\
\Delta {\cal L}^{(1)} &=&
\frac{1}{m}\left[ \xi U''(X) - 
\xi F'(X,\tau) + \overline{\xi F'(X,\tau)} \right]
\frac{\partial}{\partial V} \nonumber \\ 
&~&+ V \xi' \frac{\partial}{\partial X}
+ V \dot\xi' ~\frac{\partial}{\partial V},
\nonumber \\
\Delta {\cal L}^{(2)} &=&
\frac{1}{2m}\left[ (\xi^2 - \overline{\xi^2}) U'''(X) - 
\xi^2 F''(X,\tau)\right]
\frac{\partial}{\partial V} \nonumber \\
&~&- V 
(\xi'^2-\overline{\xi'^2} )\frac{\partial}{\partial X}
-V \dot\xi'\xi' ~\frac{\partial}{\partial V},
\end{eqnarray} 
where the {\it bar} over the terms indicates an average over a time
period.
The perturbative asymptotic solution of Eq.(\ref{pertFP}) can be
formally written as follows
\begin{equation}
\label{pertFPformal}
\tilde{P}_{\infty}(X,V,\tau) = Q_{\infty}(X,V,\tau) +
\frac{1}{ \partial_{\tau} - {\cal L}}
\Delta {\cal L} ~\tilde{P}_{\infty}(X,V,\tau),
\end{equation}
where $Q_{\infty}(X,V,\tau)$ is the 
right eigenfunction of ${\cal L}$ with eigenvalue zero. 
Since the asymptotic distribution is periodic and the nonzero
eigenvalues of ${\cal L}$ have a nonvanishing real part, it 
follows from Eq.(\ref{pertFP}) 
that  $\Delta {\cal L} \tilde{P}_{\infty}$ has no overlap with
the eigenfunction of $\partial_{\tau} - {\cal L}$ with zero
eigenvalue. Hence the operation of the inverse of 
$\partial_{\tau} - {\cal L}$
is well defined as it acts only on the space of functions 
orthogonal to that eigenfunction.

It is not possible to obtain an
explicit form for the inverse of $\partial_{\tau} - {\cal L}$ 
for arbitrary $U(X)$ and $F(X,\tau)$, in which case
numerical or variational methods might be adopted to determine the
eigenfunctions and eigenvalues of ${\cal L}$ or ${\cal L}_0$ and to obtain 
$(\partial_{\tau} - {\cal L})^{-1}$. If for some form of the
potential $U(X)$, we are able to find all the eigenfunctions of, say,
${\cal L}_0$ then this inverse can
be conveniently expanded to any order in $\xi$ as follows
\begin{equation}
\label{invope:0}
\frac{1}{ \partial_{\tau} - {\cal L}}
~=~ \sum_{n=0}^{\infty}
\left[
\frac{1}{ \partial_{\tau} - {\cal L}_0} ({\cal L} - {\cal L}_0) 
\right]^n ~\frac{1}{ \partial_{\tau} - {\cal L}_0}.
\end{equation}
But a more suitable series expansion can be given for this 
operator when either $\omega$ or $\gamma/m$ is large. If
$\omega$ is also much larger than  $\gamma/m$, then this suitable
expansion is
\begin{equation}
\label{invope:1}
\frac{1}{ \partial_{\tau} - {\cal L}}
~=~ \sum_{n=0}^{\infty}
\left(
\frac{1}{ \partial_{\tau} }~ {\cal L} 
\right)^n ~ \frac{1}{ \partial_{\tau} }.
\end{equation} 
In the case when $\gamma/m$ is comparable to $\omega$, we can
write the inverse in terms of an operator ${\cal L}_V$ containing   
terms of $O(\gamma/m)$ only and which is given as  
\begin{equation}
\label{LVoperator}
{\cal L}_{V} \equiv
\frac{\gamma}{m}\frac{\partial}{\partial V} V
+ \frac{\gamma}{\beta m^2} \frac{\partial^2 }{\partial V^2}
= e^{-\frac{\beta}{4} m V^2} 
\left( -\frac{\gamma}{m} a^\dagger a \right)
e^{\frac{\beta}{4} m V^2},
\end{equation}
where the operators $a$ and $a^\dagger$ follow the commutation relation
$[a,a^\dagger] = 1$ and are defined as
\begin{equation}
\label{aadagger}
a = \frac{1}{\sqrt{\beta m}}\frac{\partial}{\partial V}
+ \frac{\sqrt{\beta m}}{2} V,
~~
a^\dagger = -\frac{1}{\sqrt{\beta m}}\frac{\partial}{\partial V}
+ \frac{\sqrt{\beta m}}{2}V.
\end{equation}
So in this case the convenient expansion of the inverse operator is
\begin{eqnarray}
\label{invope}
\frac{1}{ \partial_{\tau} - {\cal L}}
&=& \sum_{n=0}^{\infty}
\biggl[
\frac{1}{ \partial_{\tau} - {\cal L}_V} ~({\cal L} - {\cal L}_V) 
\biggr]^n \frac{1}{ \partial_{\tau} - {\cal L}_V}
\nonumber \\
&=& e^{-\frac{\beta}{4} m V^2}
\biggl[
\sum_{n=0}^{\infty}
\biggl(
\frac{1}{ \partial_{\tau} + \frac{\gamma}{m} a^\dagger a } 
~({\cal L} - {\cal L}_V) \biggr)^n  \nonumber \\
& &  \times 
~\frac{1}{ \partial_{\tau} + \frac{\gamma}{m} a^\dagger a }
\biggr]
e^{\frac{\beta}{4} m V^2}~.
\end{eqnarray}
The idea of writing this operator in terms of $a$ and $a^\dagger$ is
that its action on $h(V)\exp(-\beta m V^2/2)$, where 
$h(V)$ is some polynomial of $V$, can be determined more easily
since it involves the action of a specific function
of $a^\dagger a$ on a series made of eigenfunctions of
$a^\dagger a$.

The calculational scheme is thus reduced to the following: Find the
right eigenfunction $Q_{\infty}$ of ${\cal L}$
with eigenvalue zero to the desired order in $\xi$. 
Then use one of the truncated series 
(\ref{invope:0}), (\ref{invope:1}), or (\ref{invope}), such that the
truncation is consistent with the chosen order, keeping in mind that 
$\xi$ depends on $\omega$. Next substitute this truncated inverse operator
in Eq.(\ref{pertFPformal}) and from it extract 
$\tilde{P}_{\infty}(X,V,\tau)$ order by order. 

\section{Asymptotic distribution: A Perturbative Analysis}
\label{asympert}

\subsection{First-order perturbation}   
\label{firstorder}
We now find the asymptotic distribution $P_{\infty}(x,v,t)$, to
first order in $\xi$, with the condition that it vanishes at the boundary
of $(x,v)$ space. We will also show that the time average of this
distribution $\overline{{P}_{\infty}(x,v,t)}$ is the canonical equilibrium
distribution with $U(x)$ replaced by $U_{eff}(x)=U(x)+U^{(1)}(x)$, where
$U^{(1)}(x)$ will be explicitly evaluated. 

To this end we need to
determine  $Q_{\infty}(X,V,\tau) = \sum_n Q_{\infty}^{(n)}(X,V,\tau)$  to
the same order from the equation ${\cal L}Q_{\infty}(X,V,\tau)=0$. As
before, the superscript on $Q_{\infty}^n$ indicates the corresponding
order
in $\xi$.
 The zeroth order $ Q_{\infty}^{(0)}$ is the solution to the
equation ${\cal L}^{(0)}Q_{\infty}^{(0)}=0$ which yields the equilibrium
distribution 
\begin{eqnarray}
\label{ptil0}
\tilde{P}_{\infty}^{(0)}(X,V,\tau)&=& Q_{\infty}^{(0)}(X,V,\tau)
\nonumber \\
&=& \frac{1}{Z_0} 
\exp\biggl(-\beta \biggl[\frac{1}{2}mV^2 + U(X)\biggr]\biggr).
\end{eqnarray} 

The first-order $\tilde{P}_{\infty}^{(1)}$ is obtained 
from Eqs.(\ref{pertFPformal}) and (\ref{invope}) 
\begin{equation}
\label{ptil1}
\tilde{P}_{\infty}^{(1)}= 
Q_{\infty}^{(1)} + e^{\frac{-\beta}{4} m V^2}
\frac{1}{ \partial_{\tau} + \frac{\gamma}{m} a^\dagger a }
e^{\frac{\beta}{4} m V^2}
 \Delta {\cal L}^{(1)} ~\tilde{P}_{\infty}^{(0)},
\end{equation}
where $ Q_{\infty}^{(1)}$ is the solution to the equation 
${\cal L}^{(0)}Q_{\infty}^{(1)} + {\cal L}^{(1)}Q_{\infty}^{(0)}=0$. 
This solution is straightforward to determine since
${\cal L}^{(1)}Q_{\infty}^{(0)}$ has the same form as the right hand
side of the identity:
${\cal L}^{(0)}[h(X)Q_{\infty}^{(0)}]= - V h'(X)Q_{\infty}^{(0)}$ for
any arbitrary function $h(X)$. 
We get
\begin{eqnarray}
\label{Q1}
Q_{\infty}^{(1)}(X,V,\tau)= -\beta U^{(1)}(X) \tilde{P}_{\infty}^{(0)}~,
\nonumber \\
\frac{\partial}{\partial X} U^{(1)}(X) =
-\overline{\xi(X,\tau) F'(X,\tau)}~.
\end{eqnarray}
In Eq.(\ref{ptil1}) we have used the inverse operator (\ref{invope}) 
truncated right after the first term which amounts to neglecting 
$O(\xi/\omega)$ terms. To this approximation, it is 
then sufficient to keep only the term proportional to 
$\dot\xi$ in $\Delta{\cal L}^{(1)}$, which is rewritten as follows
\begin{eqnarray}
\Delta {\cal L}^{(1)}&=& e^{\frac{-\beta}{4} m V^2}
\left[
-\sqrt{\frac{\beta}{m}}( \xi U'' - \xi F' + \overline{\xi F'} )
a^\dagger  \right. \nonumber \\
& &+ \left.\xi' \frac{\partial}{\partial X} \frac{a+a^\dagger}{\beta m}
-\dot\xi' (a+a^\dagger)a^\dagger
\right] e^{\frac{\beta}{4} m V^2}.
\end{eqnarray}
Hence  Eq.(\ref{ptil1}) upon neglecting terms of 
$O(\xi/\omega)$ reduces to
\begin{eqnarray}
\label{ptil1:1}
\tilde{P}_{\infty}^{(1)}(X,V,\tau) &=& 
-\beta U^{(1)}(X)\tilde{P}_{\infty}^{(0)}
- e^{\frac{-\beta}{4} m V^2}
\frac{1}{ \partial_{\tau} + \frac{\gamma}{m} a^\dagger a }
\nonumber \\
& &\times\left[  \dot\xi (a+a^\dagger) a^\dagger \right]
e^{-\frac{\beta}{4} m V^2} ~
\frac{1}{Z}e^{-\beta U(X)}
\nonumber \\
&=& -\beta
\left[ U^{(1)}(X) + K^{(1)}(V;\xi)
\right]\tilde{P}_{\infty}^{(0)}(X,V,\tau), \nonumber \\
& &
\end{eqnarray}
where
\begin{eqnarray}
\label{k1}
K^{(1)}(V;\xi)
&=&\frac{1}{\omega^2 + 4\frac{\gamma^2}{m^2}}
\left[ \frac{2\gamma}{\beta m}
\left(\frac{2\gamma}{m}\xi'- \dot\xi'\right) \right. \nonumber \\
& &+ \left.
\left(\omega^2 \xi' + \frac{2\gamma}{m}\dot\xi'\right) mV^2
\right],
\end{eqnarray}
which simplifies to $K^{(1)}(V;\xi) \approx \xi' m V^2$ 
when $\omega \gg \gamma/m$.

We now get the asymptotic distribution
from Eqs. (\ref{coord}), (\ref{ptil0}), and (\ref{ptil1:1}),
\begin{eqnarray}
\label{pdist1}
{P}_{\infty}(x,v,t) &=&\frac{1}{Z}
\exp\biggl(-\beta \biggl[\frac{1}{2}m v^2 + U(x) + U^{(1)}(x) 
\nonumber \\ 
& &+ K^{(1)}(v;\xi) - \xi U'(x) - \dot\xi mv \biggr]\biggr).
\end{eqnarray}
Unlike in the distributions for static potentials,  
here the velocity $v$ gets coupled to the position $x$ through $\xi$.
The averaged large-time distribution is given as 
\begin{equation}
\label{pavgdist1}
\overline{{P}_{\infty}(x,v,t)} = 
\frac{1}{Z} \exp\biggl(-\beta 
\biggl[ \frac{1}{2}m v^2 + U(x) + U^{(1)}(x)\biggr]\biggr),
\end{equation}
where the explicit expression for $U^{(1)}(x)$ is obtained upon
substituting $\xi$ in Eq.(\ref{Q1}) and then integrating: 
\begin{eqnarray}
\label{U1}
U^{(1)}(x)&=& \frac{1}{4m(\omega^2 + \frac{\gamma^2}{m^2})}
\biggl[ \biggl( f^2(x)+ g^2(x) \biggr)  \nonumber \\
& &+ \frac{2\gamma}{m\omega}\int^x dy 
\biggl( g(y)f'(y) - f(y)g'(y)\biggr)
\biggr].
\end{eqnarray}
Thus the time-averaged large-time behavior of the Brownian particles is
described by the canonical distribution at a temperature $\beta^{-1}$ with
an effective potential $U_{eff}=U+U^{(1)}$
that depends on the frequency and space dependence of the driving force in
addition to the properties of the particle. Note that a nontrivial
contribution to the effective potential arises to this order only if $f$
or $g$ are space dependent, and that it can be tuned by varying $f$, $g$,
or $\omega$.

The additional term $U^{(1)}$ is the average energy associated with the
rapid motion. This can be seen easily upon substituting for $F$ in
terms of $\xi$ using Eq. (\ref{xieqn}) and rewriting $U^{(1)}$ as follows:
\begin{eqnarray}
\label{uinterpret}
U^{(1)}(x) &=&
-\int^x dy~\overline{[\xi(y,t) F'(y,t)]} 
\nonumber \\
&=& \frac{1}{2} m 
\overline{\bigl[\dot\xi(x,t)\bigr]^2}
- \int^x dy \overline{\left(-\gamma \dot\xi(y,t) 
\frac{\partial}{\partial y}\xi(y,t)\right)}.
\end{eqnarray}
Thus $U^{(1)}$ is sum of the average kinetic energy and the work done
against the damping on the fast variable $\xi$.

A first-order perturbation treatment is justified, provided
${\cal L}^{(2)}$ and $\Delta {\cal L}^{(2)}$
are negligible when compared to 
${\cal L}^{(1)}$ and $\Delta {\cal L}^{(1)}$, respectively.
The criterion for this is
$| \xi{\partial f}/{\partial x}| \gg 
| U'''{\xi}^2|$. For instance, suppose that the length scales
over which $U(x)$ and $F(x,t)$ vary are comparable and $\alpha \omega_0$
is a typical frequency associated with anharmonic terms. Then the above
criterion  reduces to 
$(\omega^2 + \gamma^2/m^2) \gg \alpha^2\omega_0^2$, 
which is consistent with $\xi$ being small.

It should be remarked that the restriction of periodicity of $F(x,t)$ to
the fundamental frequency $\omega$ is not essential; one can include the
higher harmonics as well. Also by considering 
generalizations to higher dimensions or to many interacting
particles one does not encounter any additional
computational difficulties in the evaluation of the distribution. The
only additional assumption needed to write down $U_{eff}$ is that
$\vec{f}(\vec{x})$ and $\vec{g}(\vec{x})$ are curl free.

\subsection{Second-order perturbation}   
\label{secondorder}

We now calculate the second-order corrections to the asymptotic
distribution. Further, in this subsection, we restrict our treatment
to frequencies that satisfy the condition 
$\omega \gg \gamma/m$, in which case the first-order term 
$\tilde{P}_{\infty}^{(1)}$ is as given in Eq.(\ref{ptil1:1})  
with $K^{(1)}(V;\xi) = \xi' m V^2$. The second-order term of 
Eq.(\ref{pertFPformal}) is
\begin{eqnarray}
\label{ptil2}
\tilde{P}^{(2)}_{\infty} 
&=& Q^{(2)}_{\infty} + 
\frac{1}{ \partial_{\tau} - {\cal L}}
[\Delta {\cal L}^{(1)} \tilde{P}^{(0)}_{\infty}
+\Delta {\cal L}^{(1)} \tilde{P}^{(1)}_{\infty}
+\Delta {\cal L}^{(2)} \tilde{P}^{(0)}_{\infty}]
\nonumber \\
&=& Q^{(2)}_{\infty} +
\left\|\frac{1}{ \partial_{\tau} - {\cal L}^{(0)}}
\Delta {\cal L}^{(1)} \tilde{P}^{(0)}_{\infty}~\right\|
+\frac{1}{ \partial_{\tau}}
\Delta {\cal L}^{(1)} \tilde{P}^{(1)}_{\infty}
\nonumber \\
& &+ \frac{1}{ \partial_{\tau}}
\Delta {\cal L}^{(2)} ~\tilde{P}^{(0)}_{\infty} + 
O(\frac{1}{\omega}\xi^2).
\end{eqnarray}
Though the second term in the above expression is of $O(\xi)$
it has been included here because when calculated earlier, to first order 
in $\xi$, the terms of $O(\xi/\omega)$ were neglected. 
This term is written within {\it pipes}
to indicate that only terms of $O(\xi/\omega)$ and
$O(\xi/\omega^2)$ have to be retained and not the
$O(\xi/\omega^0)$ term that has already been included in
$\tilde{P}^{(1)}_{\infty}$.

We now evaluate the terms on the right-hand side of 
Eq.(\ref{ptil2}). The first term 
$ Q_{\infty}^{(2)}$ is the solution to the equation 
${\cal L}^{(0)}Q_{\infty}^{(2)}  
+{\cal L}^{(1)}Q_{\infty}^{(1)}
+{\cal L}^{(2)}Q_{\infty}^{(0)}
=0$. 
This, as in the case of $Q_{\infty}^{(1)}$,
 is straightforward to determine and we get
\begin{eqnarray}
\label{Q2}
Q_{\infty}^{(2)}(X,V,\tau)&=& 
-\beta {\cal U}_1(X) \tilde{P}_{\infty}^{(0)}
+ \frac{\beta^2}{2} \left(U^{(1)}(X)\right)^2 
\tilde{P}_{\infty}^{(0)},
\nonumber \\
\frac{\partial}{\partial X} {\cal U}_1(X) &=&
-\overline{(\xi')^2}~U'(X)  
+\frac{1}{2}\overline{(\xi)^2}~U'''(X).
\end{eqnarray}
The second term is
\begin{eqnarray}
\label{term2}
& &\left\|\frac{1}{ \partial_{\tau} - {\cal L}^{(0)}}
\Delta {\cal L}^{(1)} ~\tilde{P}^{(0)}_{\infty}~\right\|
\nonumber \\
& & = \left\|\left( 
\frac{1}{ \partial_{\tau}} 
+\frac{1}{ \partial^2_{\tau}}  {\cal L}^{(0)}
+\frac{1}{ \partial^3_{\tau}}  ({\cal L}^{(0)})^2
+\cdots \right)
\Delta {\cal L}^{(1)} ~\tilde{P}^{(0)}_{\infty}\right\|
\nonumber \\
& &=-\beta {\cal K}(X,V;\xi)~\tilde{P}^{(0)}_{\infty},
\end{eqnarray}
where 
\begin{eqnarray}
\label{Kdef}
{\cal K}(X,V;\xi)~\tilde{P}^{(0)}_{\infty}&=&
-\frac{\partial_{\tau}}{\omega^2}~H(X,V;\xi)
-\frac{1}{\omega^2}{\cal L}^{(0)}~H(X,V;\xi),
\nonumber \\
H(X,V;\xi)&=& 
\left( \xi U'' +\xi'U'- \frac{1}{4} 
\xi F' + \frac{1}{4}\overline{\xi F'} \right)
V \tilde{P}^{(0)}_{\infty}  \nonumber \\
& &+ {\cal L}^{(0)}\left(mV^2 \xi' \tilde{P}^{(0)}_{\infty} \right).
\end{eqnarray}
The third and the fourth terms are
\begin{eqnarray}
\label{term34}
\frac{1}{ \partial_{\tau}}
\Delta {\cal L}^{(1)}~\tilde{P}^{(1)}_{\infty}
&=& -\beta m V^2 [(\xi')^2 -\overline{(\xi')^2}]
\tilde{P}_{\infty}^{(0)} \nonumber \\
& &+ \beta^2 [
U^{(1)}mV^2\xi' + \frac{1}{2}(mV^2\xi')^2] \tilde{P}_{\infty}^{(0)},
\nonumber \\
\frac{1}{ \partial_{\tau}}
\Delta {\cal L}^{(2)} ~\tilde{P}^{(0)}_{\infty}
&=& \frac{1}{2}
\beta m V^2 [(\xi')^2 -\overline{(\xi')^2}]
\tilde{P}_{\infty}^{(0)} \nonumber \\
& &-\beta [{\cal U}_2(X) + C(X,V)]\tilde{P}_{\infty}^{(0)},
\end{eqnarray}
where 
$\left({\cal U}_2\right)' 
= \overline{\xi\xi'} U''- 3 \overline {\xi'\xi'} U'$ and 
$C(X,V)$ is an arbitrary $\tau$-independent function. These 
arbitrary and $\tau$-independent terms are included since the action of
$\partial_{\tau}^{-1}$ allows this ambiguity. This ambiguity is removed
from the condition obtained by substituting $\tilde{P}_{\infty}$ in
Eq.(\ref{pertFP}) and averaging over a period. Thus 
$\tilde{P}^{(2)}_{\infty}$ should satisfy
\begin{equation}
\label{avgp2}
 {\cal L}^{(0)}\overline{\tilde{P}^{(2)}_{\infty}}
+ \overline{\Delta {\cal L}^{(1)}\tilde{P}^{(1)}_{\infty}} =0, 
\end{equation}
which implies, after some calculation, that $C$ is the solution of the
equation 
\begin{equation}
\label{avgp2:2}
{\cal L}^{(0)} [C\tilde{P}_{\infty}^{(0)}] 
+ 4 m \overline{\xi'\xi''} V^3 \tilde{P}_{\infty}^{(0)}=0.
\end{equation}
In the high-viscous limit, $\gamma/m \gg \omega_0$, this function   
$C(X,V)\approx(4m/3\gamma) \overline{\xi'\xi''}
\left(m V^3+6V/\beta\right)$.
Using Equations (\ref{ptil2})-(\ref{term34}) and 
Eq.(\ref{ptil1:1}) we get the asymptotic distribution to second order
\begin{eqnarray}
\label{pdist2}
{P}_{\infty}(x,v,t) 
&=& \frac{1}{Z}
\exp\biggl(-\beta \biggl[\frac{1}{2}m V^2
+ U(X) + U^{(1)}(X) \nonumber \\
& &+U^{(2)}(X) +R(X,V,t)\biggr] \biggr)
\end{eqnarray}
where 
$ R(X,V,t)=  K^{(1)}(V;\xi) + {\cal K}(X,V;\xi)
+\frac{1}{2}mV^2[(\xi'(X,t))^2-\overline{(\xi'(X,t))^2}] +C(X,V)$;
$U^{(2)}(X) = {\cal U}_1(X)+ {\cal U}_2(X)$;
and, to second order,
$X=x-\xi(x,t)-\xi(x,t)\xi'(x,t)$ and
$V=v-\dot\xi(x,t)-\xi(x,t)\dot\xi'(x,t)$.
It is clear from the above equation that  
$\overline{{P}_{\infty}(x,v,t)}$ will contain terms with both $v$ and
$x$ dependence in addition to purely $x$-dependent and $v$-dependent
terms. 

The explicit form of $U^{(2)}(X)=U^{(2)}(x)$ which contributes at
second order to the effective static potential is
\begin{eqnarray}
\label{U2}
U^{(2)}(x)&=&  
\frac{1}{4m^2\omega^2(\omega^2 + \frac{\gamma^2}{m^2})}
\biggl[ 
\biggl( f(x)^2+ g(x)^2 \biggr) U''(x) \nonumber \\
& &-8 \int^x dy \biggl( \bigl(f'(y)\bigr)^2 + \bigl(g'(y)\bigr)^2 \biggr) 
U'(y) \biggr].
\end{eqnarray}
We notice here that a nontrivial contribution arises even for
$x$-independent driving provided $U(x)$ is anharmonic. Also note that 
this second-order correction to the effective potential depends on
$U(x)$ while this is not the case at the first order.

We recover the results of Devoret {\it et. al.} \cite{devoret} and Jung
\cite{jung} when $f(x)$ and $g(x)$ are independent of the position
coordinate $x$. In the former reference the inertial term was considered
while in the latter it was not, but their analysis of the high-frequency
limit is tantamount to assuming that the driving
frequency is larger than all typical frequencies of the
system including that of noise. This assumption restricts the validity of
the answer, and clearly does not hold for white noise. In fact, this
assumption would lead to the absence of the term 
$\int^x [(f')^2 +(g')^2 ] U'$ in the effective
potential for $x$-dependent driving. That the error, when this
assumption is made,shows up at $O(\xi^2)$ is also evident from the
presence of the term $V \overline{\xi'^2}\partial/\partial X$ 
in ${\cal L}^{(2)}$ operator.

We reiterate that there is a nontrivial contribution to the effective
potential from the position-dependent driving forces. First, it shows up
at first order itself and, second, it shows up at second order
even for harmonic potentials; both of which are absent for
position-independent driving forces.

\section{Slowly Oscillating forces}    
\label{slowfields}
In this section, we will consider the opposite extreme, namely, 
$\omega$ small compared to both $\gamma/m$ and $\omega_0$. The aim is to
compare with the asymptotic behavior we found in the previous
section under rapid driving. Under slow driving, the Brownian particle
sees an unchanging potential
within its relaxation time and so the Boltzmann-like
distribution corresponding to the instantaneous potential is
a good zeroth-order starting point for perturbation.

The large-time distribution $P_{\infty}(x,v,t)$ will acquire
the same periodicity as that of the driving force. Hence
the left-hand side of Eq.(\ref{FPeqn}) is of $O(\omega)$ while the
right-hand side has two terms, one of $O(\gamma/m)$ and the other
of $O(\omega_0)$. Thus $P_{\infty}(x,v,t)$ can be evaluated
perturbatively in $(m\omega/\gamma)$ and $(\omega/\omega_0)$. To
the leading (zeroth) order, this distribution is 
\begin{eqnarray}
\label{zerokramers}
P_{\infty}^{(0)}(x,v,t) &=& \frac{1}{Z(t)}
\exp\{-\beta [ \frac{1}{2}mv^2 + U(x) 
\nonumber \\
& & + U_f(x)\cos(\omega t)+ U_g(x)\sin(\omega t))] \},
\end{eqnarray}
where $U_f(x)=-\int^x dy f(y)$ and $U_g(x)=-\int^x dy g(y)~$;
$Z(t)$ is the instantaneous normalisation constant determined 
by the normalization condition: $\int dx P_{\infty}(x,v,t) = 1$ at any
time $t$, and evidently satisfies the periodicity condition $Z(t+T)=Z(t)$.

We now estimate the averaged large-time distribution 
$\overline{P_{\infty}(x,v,t)}$ at low and high temperatures separately.
To determine the average distribution at high temperature, it is
convenient to rewrite this equation as follows. 
\begin{eqnarray}
\label{solnapprox}
P_{\infty}(x,v,t) 
&=& \frac{1}{Z(t)}
e^{-\beta [\frac{1}{2}mv^2 + U(x)] }~ 
{\cal I}_0 \left( \beta V(x)\right) \nonumber \\
& &\times \left
[1 + \sum_{n \neq 0} e^{{\it i} n \omega t} 
\frac
{{\cal I}_n\left( \beta V(x)  \right)}
{{\cal I}_0\left( \beta V(x)  \right)}
\right],
\end{eqnarray}    
where ${\cal I}_n(\alpha)$ is the modified Bessel function
and $V(x) = \sqrt{[U_f(x)]^2 +[U_g(x)]^2} $. The ratio
${{\cal I}_n(\alpha)}/{{\cal I}_0(\alpha)}$ lies between $0$ and $1$
for $0 \leq \alpha < \infty$, and decreases very rapidly for small
$\alpha$;
$ {{\cal I}_n(\alpha)}/{{\cal I}_0(\alpha)}
\sim {\alpha^n}/{2^n n!}$.
Thus it suffices to keep a few terms, and to leading order we obtain 
$\overline{P_{\infty}(x,v,t)} 
=(1/Z) 
\exp\{-\beta [\frac{1}{2}mv^2 + U_{eff}(x)]\}$, 
where
\begin{equation}
\label{slowueff}
U_{eff}(x) = U(x) - \frac{1}{\beta}
\ln \{ {\cal I}_0
( \beta \sqrt{[U_f(x)]^2 +[U_g(x)]^2} ) \}.
\end{equation}
Thus the Brownian particles, when observed over a time of $O(T)$,
get described by the canonical distribution at a temperature $\beta^{-1}$
with an effective potential that depends on the temperature. 

At low temperature the saddle-point approximation can be used to
evaluate $\overline{P_{\infty}(x,v,t)}$. To the leading order this will be
$
\overline{P_{\infty}(x,v,t)}=P_{eq}(v)~ 
\overline{[1/N(t)]\sum^{N(t)}_{i=1}
\delta(x-x^{(i)}_{min}(t))}$,
where $\{x^{(i)}_{min}(t)\}$ are the minima at time $t$ of the
function $U(x) + U_f(x)\cos(\omega t) + U_g(x)\sin(\omega t)$ and $N(t)$
is the number of minima at that time.

It might appear that $U_f(x)$ and $U_g(x)$, obtained by
integrating $f(x)$ and $g(x)$, respectively, have 
different constants of integration in regions where the driving forces are
present from those where they are absent. However, they get fixed by the
condition that $P(x,v,t)$ and $\partial P(x,v,t)/\partial x $
are single-valued functions of $x$.

\section{Discussion}
\label{concl}
In this section, we review the results of this paper for a general
position-dependent periodic driving force with a focus on the high
frequency regime. We then discuss effective potentials
for specific forms of the driving functions, and use these results to
point out some possible applications. The nature of these effective
potentials can be drastically different from that of the original
potentials, and we illustrate this fact using the example of a simple
pendulum.

The principal results of this paper concern the form of the asymptotic
distribution of a
Brownian particle subjected to a driving force, which is periodic in time
but is an arbitrary function of position. In the limit of high-frequency
driving, the particle makes small, rapid excursions around a smooth path
along which the motion is relatively slow. This affords the possibility of 
a systematic perturbative treatment in powers of the excursion amplitude.
The result for the distribution averaged over a cycle is described in
terms of an effective potential whose form we derived. Interestingly, the
leading contribution $U^{(1)}$, which is second order in the amplitude of
the applied driving force, is present only if the driving force is
position dependent. This $U^{(1)}$ can be interpreted as the average
energy of the excursion variable. In the next-order contribution to the
effective potential as well, this position dependence of the driving force
has an interesting effect even for purely harmonic
confining potentials. 

Central to our discussion of rapid periodic driving is the
separation into slow and fast variables. It should be noted that this
demarcation of slow and fast is based on whether or not the variable
varies considerably over a time period. This is different from the
distinction made in discussions of fast and slow variables that are so
categorized according to whether they relax in small or large times. In
this latter case, many methods have been developed to eliminate the fast
variables and obtain an effective dynamics for the relaxation of the slow
variables\cite{riskenkampen,grigolini}. 

The effective potential that we find can be qualitatively different in the
low- and high-frequency regimes.  
When $\omega$ is small, the leading additional potential felt by the
particles, $\Delta U(x) \equiv U_{eff}(x)-U(x)$, is always nonpositive
for any choice of $f(x)$ and $g(x)$. On the other hand  for large
$\omega$, $\Delta U(x)$ can be either positive or negative in general,
though if one of $f(x)$ or $g(x)$ is identically zero then it is always
non-negative. Also to the leading order when $\omega$ is
small, $\Delta U(x)$ depends on temperature but not on $\omega$, whereas
in large $\omega$ limit it only depends on $\omega$ and not on
temperature.

Certain choices of the driving force can lead to interesting outcomes.
Suppose we choose $U(x)$ to be a confining potential which is 
a monotonically increasing function of $|x|$ while $f(x)$ is a
monotonically decreasing function of $|x|$ which vanishes at $|x|=L$, and
$g(x)$ vanishes identically:
for example, $U(x)= \frac{1}{2} m \omega_0^2 x^2$ and
$f(x)=f_0 \sin(\pi x/L)$ for $-L \le x \le L$ and zerootherwise.
When $\omega$ is small the effective potential continues to have a single
minimum at the origin. But when $\omega$ is large, it develops two
additional minima; one close to $L$, between $0$ and $L$ and the other
close to $-L$, between $-L$ and $0$. Thus, a system of Brownian
particles would cluster in a region near the origin for low driving
frequency, while at high frequency these particles would segregate into
two clusters which are separated by a distance of $O(L)$.
 
The parameters specifying the particles also enter the effective
potential. Hence the minima of the effective potential as seen by
different species of particles are different.
One can make use of this fact, for instance, to separate different 
species of Brownian particles in a situation where they
are initially mixed, by driving them with a space-dependent periodic
force.
 
To get an idea of the magnitude of the qualitative change the effective
potential introduces it is instructive to examine an example. 
Consider a rigid pendulum: a massless rod of length $l$ with a bob of mass
$m$ attached to it at the end and oscillating in a gravitational 
field $g$. The
potential has extrema at $\theta = 0$  and $\theta =\pi$ which are
respectively stable and unstable points, where $\theta$ is the angle the
rod makes with the negative $y$ axis (along the direction of
gravity). Now oscillate the point of suspension along the $y$ axis with a
frequency $\omega$ and amplitude $a$. The angle $\theta$ then evolves
according to the equation 
$l \ddot{\theta} + g \sin\theta = a \omega^2 \sin\theta \cos(\omega t)$.
The effective potential
$V_{eff}(\theta)$ has extrema at $\theta =0,\pi$, and $\theta_{\pm}$,
where 
$\theta_{\pm}= \pi \pm \cos^{-1}\lambda$ with  
$\lambda = 2gl/a^2\omega^2$. The stability of these points is as follows: 
$\theta =0$ is stable; $\theta =\theta_{\pm}$ exist only if 
$\lambda < 1$ and when they exist they are unstable; $\theta =\pi$ is
stable when $\theta_{\pm}$ exist and is unstable otherwise. In a nutshell,
when $\lambda \geq 1$ the pendulum shows no qualitative change in its
behavior upon oscillating the point of suspension whereas when 
$\lambda < 1$ then $\theta = \pi$ also becomes a stable
point and hence the pendulum can make small oscillations about this point
too. This dramatic change in the behavior of the pendulum was
experimentally demonstrated by Kapitza\cite{kapitza}.  

Interestingly, if we oscillate the point of suspension along the
$x$ axis instead of the $y$ axis, the pendulum will exhibit
a different behaviour. In this case, the effective potential has extrema
at $\theta =0,\pi$, and $\pm \cos^{-1}\lambda$ if $\lambda < 1$; the
points $\pm \cos^{-1}\lambda$ are stable when they exist; 
$\theta =0$ is unstable if $\lambda < 1$ and stable otherwise;
$\theta =\pi$ is always unstable. In other words, the pendulum now
oscillates about points that do not lie on the $y$ axis. The nature of the
extrema of the effective potential does not change in the presence of
the viscous term and noise except that the value of $\lambda$ will now be
different (with $\omega^2$ replaced by 
$\omega^4 /( \omega^2 + ({\gamma}/{m})^2 )$). 

To conclude, we have developed a perturbative calculational scheme 
to study the asymptotic behavior of Brownian particles under the
influence of rapidly oscillating forces. When these forces are
position dependent, nontrivial effects are seen in the large-time
behavior. The formalism developed here can be generalized
straighforwardly to interacting Brownian particles in any dimensions. It
can also be extended to study the behavior of fluctuating fields when
subjected to periodic driving.

\acknowledgments
The authors would like to thank Deepak Dhar for helpful comments and
discussions. MB also acknowledges useful conversations with Ashutosh
Sharma.



\end{multicols}
\end{document}